\documentclass[twocolumn, showpacs, preprintnumbers, dvips,amsmath, amssymb]{revtex4}

\usepackage{mathrsfs}
\usepackage{amsfonts}
\usepackage{graphicx}
\usepackage{dcolumn}
\usepackage{bm}

\begin{document}

\preprint{...}

\title{Anyonic Loops in Three Dimensional Spin liquid and Chiral Spin Liquid}

\author{Tieyan Si}\email{sity@itp.ac.cn}\author{Yue Yu}
\affiliation{Institute of Theoretical Physics, Chinese Academy of
Sciences, P.O. Box 2735, Beijing 100080, China}

\date{\today}

\begin{abstract}

We established a large class of exactly soluble spin liquids and
chiral spin liquids on three dimensional helix lattices by
introducing Kitaev-type's spin coupling. In the chiral spin
liquids, exact stable ground states with spontaneous breaking of
the time reversal symmetry are found. The fractionalized loop
excitations in both the spin and chiral spin liquids obey
non-abelian statistics. We characterize this kind of statistics by
non-abelian Berry phase and quantum algebra relation. The
topological correlation of loops is independent of local order
parameter and it measures the intrinsic global quantum
entanglement of degenerate ground states.

\end{abstract}


\maketitle

\section{Introduction}

Topological order of condensed matter system is the cornerstone of
topological quantum computation\cite{das}. The topologically ordered system 
have degenerated ground states which depend on the topology instead of symmetry. 
This degeneracy of ground state is robust against any local perturbation 
due to the gap between ground state and excited states\cite{wen}. 
An interesting application of topological stablizer quantum codes is implemented 
in Kitaev's toric code model\cite{toric}. A topological color codes with 
quantum error-correcting capabilities is developed later\cite{color}, this model removes the 
need for selective addressing.

There are different ways to perform quantum computation in use of topologically ordered states. 
Usually the quantum gate or unitary transformation are implemented by brading quasiparticles. 
The transversal implementation of quantum gate provides us another way of 
topological quantum computation without braiding\cite{nobraid}, in which the stablizer color codes is achieved 
in a 3-dimensional lattice model. While we focus on the braiding of loop quasiparticles in 
3-dimensional spin liquid and chiral spin liuid in this paper.

We established the 3-dimenisonal spin liquids by generalizing 
Kitaev's two dimensional spin model. Kitaev's spin model on honeycomb lattice has a highly 
degenerate ground state\cite{kitaev}. Abelian anyons and non-abelian anyons emerge in the background of short range spin liquid.
These anyonic excitations obey exotic statistics. Quantum
computation is based on anyon braiding. During the braiding
operation, anyons interact with one another only by encircling a
topological nontrivial cycle. A proposal of the experimental
operation of anyon braiding in Kitaev honeycomb model has been
provided by use of ultracold atoms trapped in optical
lattices\cite{cwzhang}, however there is some controversal 
on the details of this experimental proposal\cite{08014620}.

Another most promising candidate of non-abelian anyons may be
realized in fractional quantum Hall states with
$\nu=5/2$\cite{read, xia}. The fractional quantum Hall states
break parity and time-reversal symmetry since fractional quantum
Hall occurs in the presence of a strong magnetic field. Quantum
information is stored in different fusion channels of non-abelian
anyons\cite{das}. For two $\sigma$ quasiparticles following fusion
rule $\sigma\cdot\sigma\sim{\textbf{1}+\psi}$, if fusion outcome
is $\textbf{1}$, we define the qubit as state $|0\rangle$. When
fusion outcome is $\psi$, the state of the qubit is $|1\rangle$.
We may use $2n$ antidots at which quasiholes are pinned to
construct $n$ qubits.

Non-abelian anyons also arise from other two dimensional systems,
such as $Sr_{2}RuO_{4}$. In this $p+ip$ superconductor, the
half-quantum vortices with flux $hc/2e$ exhibit non-abelian
statistics\cite{das2}. Spin-$1/2$ chiral spin liquid states
proposed by Kalmeyer and Laughlin possess fractionalized
excitations\cite{kalmeyer}. This kind of non-abelian anyons
appears in 2D solvable spin models which are generalization of
Kitaev honeycomb model\cite{yao}\cite{yangsun}. These anyons are
point quasiparticles confined in two dimensions. The main
disadvantage of two dimensional system is that the expectation
value of topological symmetry operators vanishes at finite
temperature. This makes the topological memory thermally
fragile\cite{alicki}\cite{ortiz}.

Since most physical materials in nature are three dimensional,
anyonic excitations most likely exist in three dimensions. One
clue comes from a three dimensional exactly solvable spin model
constructed by generalizing Kitaev's toric code to cubic
lattice\cite{hamma}, in which the topological order of its ground
state is characterized by string net condensation as well as
membrane condensation. The string net and membrane condensation
also appear in another exactly solvable spin model\cite{brane},
which is constructed on a general 3-complex embedded in a closed
and connected base manifold in three dimension. Its ground state
degeneracy depends on the homology of the 3-manifold.

To establish three dimensional chiral spin liquid, we start from a
different way to search for anyons in three dimensions by
generalizing Kitaev honeycomb lattice model. Our interest focuses
mainly on nontrivial loop excitations. Loops confined in two
dimensions have no non-trivial entanglement without cutting each
other. It is in three dimensions that loops demonstrate much
interesting statistical phenomena without breaking topological
constraint. There is a loop braiding group for the statistics of
unknotted, unlinked closed strings\cite{baez}. For links
consisting of many entangled knots, their non-abelian statistics
is closely related to the monodromy matrix in conformal field
theory \cite{witten,ms}. Quantum loop gas applied to topological
quantum computation is under rapid developing\cite{loops}.

There were some theoretical investigations on three dimensional
chiral spin liquid based on effective field theory\cite{libby}.
However, no exact three dimensional chiral spin liquid is found so
far to the authors' knowledge. We established a large class of
exactly solvable three dimensional spin liquid and chiral spin
liquid in this paper. In these models, there is only short-range
interaction between neighboring spins. The stable ground states of
chiral spin liquid spontaneously break the time reversal symmetry.
There exist loop excitations which may serve as basic qubits to
store quantum information. Loop condensation occurs naturally at
ground state.

The projection of these three dimensional spin models to $X-Y$
plane leads to soluble two dimensional models\cite{yangsun} which
are generalization of Kitaev honeycomb model. This projection
operation does not keep time reversal symmetry. For example, the
exact spin liquid we established on 'Cu'-sublattice of crystal
green dioptase, which is the crystal structure of a material
${C{u}}_{6}{S{i}}_{6}{O}_{3}\cdot6{H_{2}O}$\cite{gros}, projects
out the very two dimensional chiral spin model constructed in Ref.
\cite{yao}.

Mandal and Surendran obtained a different exactly solvable Kitaev
model in three dimensions\cite{mandal} when we finished the
research on the spin models in this paper. None of the exact spin
models in this paper is identical to Mandal and Surendran's model.

\begin{figure}
\begin{center}\label{honey}
\includegraphics[width=0.45\textwidth]{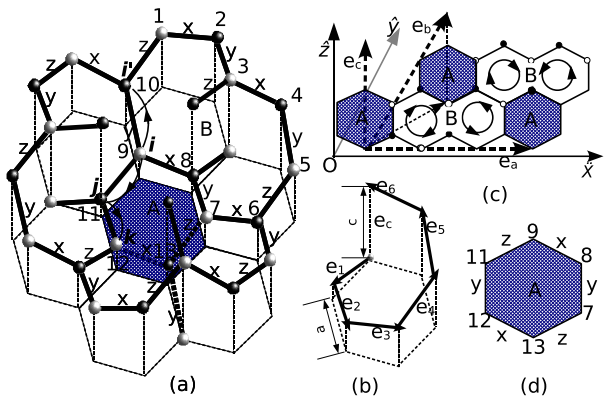}
\caption{(a) The honeycomb helix lattice. (b) A helix spin chain.
(c) The projection to $X-Y$ plane. There are two types of
elementary plaquettes, plaquette A and plaquette B . Plaquette $A$
is the shadowed hexagon. Plaquette B covers an area of double
hexagon, the oriented circles in the blank double hexagon denote
the helix chains with different chirality. (d) The bonding
configuration of the hexagon plaquette.}
\end{center}
\end{figure}

The paper is organized as follows: in section II, we constructed
exactly solvable three dimensional spin models on helix-decorated
lattice. Their ground states are invariant under time reversal
transformation. A topological phase transition from gapless phase
to gapped phase is observed. In section III, we established three
dimensional chiral spin liquids by doping spins in the helix
lattice. Their exact ground states break the time reversal
symmetry. In section IV, the topological correlation among loops
is studied from the point of view of topological quantum field
theory and Jones polynomial. In section V, non-abelian Berry phase
of loops is presented. We studied the quantum algebra of loop
statistics.

\section{Exact spin liquid on three dimensional helix
lattice}

We study the exactly soluble spin models. The exact ground state
is short-range spin liquid. They all contain helix structure, so
we classify them by the typical helix spin chain embedded in the
three dimensional lattice.

\subsection{The Honeycomb Helix lattice model}

The first spin liquid model we proposed is established on coupled
hexagonal helix lattice whose projection to the $X-Y$ plane is
Kitaev model on two dimensional Honeycomb lattice(Fig. 1). Two
neighboring hexagonal helices have opposite chirality, three pairs
of helix string surround one hexagon plaquette. The translational
invariant Hamiltonian reads
\begin{eqnarray}\label{Hexagon-helix}
H_{h}&=&-\sum_{\textbf{i}}
\{J_{z}\sigma^{z}_{\textbf{i}}\sigma^{z}_{\textbf{i}-\textbf{e}_{4}}
+J_{y}\sigma^{y}_{\textbf{i}}\sigma^{y}_{\textbf{i}+\textbf{e}_{5}}
+J_{x}\sigma^{x}_{\textbf{i}}\sigma^{x}_{\textbf{i}-\textbf{e}_{6}}\nonumber\\
&+&J_{y}\sigma^{y}_{\textbf{i}-\textbf{e}_{6}}\sigma^{y}_{\textbf{i}-\textbf{e}_{6}+\textbf{e}_{2}}
+J_{x}\sigma^{x}_{\textbf{i}-\textbf{e}_{6}+\textbf{e}_{2}}\sigma^{x}_{\textbf{i}
-\textbf{e}_{6}+\textbf{e}_{2}+\textbf{e}_{3}}\nonumber\\
&+&J_{z}\sigma^{z}_{\textbf{i}-\textbf{e}_{6}+\textbf{e}_{2}-\textbf{e}_{4}}\sigma^{z}_{\textbf{i}
-\textbf{e}_{6}+\textbf{e}_{2}}\nonumber\\
&+&J_{y}\sigma^{y}_{\textbf{i}-\textbf{e}_{6}+2\textbf{e}_{2}+\textbf{e}_{3}}
\sigma^{y}_{\textbf{i}-\textbf{e}_{6}+\textbf{e}_{2}+\textbf{e}_{3}}\nonumber\\
&+&J_{z}\sigma^{z}_{\textbf{i}-\textbf{e}_{6}+2\textbf{e}_{2}+\textbf{e}_{3}+\textbf{e}_{1}}
\sigma^{z}_{\textbf{i}-\textbf{e}_{6}+\textbf{e}_{2}+\textbf{e}_{3}}\nonumber\\
&+&J_{x}\sigma^{x}_{\textbf{i}-2\textbf{e}_{6}+2\textbf{e}_{2}+\textbf{e}_{3}}
\sigma^{x}_{\textbf{i}-\textbf{e}_{6}+2\textbf{e}_{2}+\textbf{e}_{3}}\},
\end{eqnarray}
where the sum runs over
$\textbf{i}=m\textbf{e}_{a}+n\textbf{e}_{b}+l\textbf{e}_{c}$ with
basis vector $\textbf{e}_{a}=(3\sqrt{3}a,0,0)$,
$\textbf{e}_{b}=(\frac{3\sqrt{3}}{2}a,\frac{3}{2}a,\frac{1}{3}c)$
and $\textbf{e}_{c}=(0,0,c)$(Fig. 1).($a$ and $c$ are two
different lattice spacings.) The Hamiltonian describes three
dimensional coupled hexagonal helix chains. Each one chiral
hexagonal chain is equivalent to an exactly soluble one
dimensional alternative coupling spin chain
\begin{eqnarray}\label{chiral-chain}
H_{c}=\sum_{{2j}}J_{x}\sigma^{x}_{2j}\sigma^{x}_{2j+1}+J_{y}\sigma^{y}_{2j+1}\sigma^{y}_{2j+2}+J_{z}\sigma^{z}_{2j+2}\sigma^{z}_{2j+3}.
\end{eqnarray}
The chirality of helix chain embeds in the operator product of the
three component of spin. The chiral operator
$\sigma^{z}\sigma^{y}\sigma^{x}$ marks the right-hand helix and
$\sigma^{z}\sigma^{x}\sigma^{y}$ characterize the left hand
helix(Fig. 1). Usually the non-vanishing chiral order parameter
for spin liquid state is defined as
\begin{equation}
\hat{O}_{ijk}=\sum{\vec{\sigma}_{\textbf{i}}\cdot(\vec{\sigma}_{\textbf{j}}\times\vec{\sigma}_{\textbf{k}})}.
\end{equation}
Another representation of the chiral operator in terms of the
conventional spin-1/2 fermions\cite{wen-wilczek-zee} is
\begin{equation}
\hat{O}_{ijk}=t_{ij}t_{jk}t_{ki}-t_{ik}t_{kj}t_{ji},
\end{equation}
where $t_{ij}\equiv{c^{\dag}_{i,s}c_{j,s}}$ as the hopping
operator between two lattice sites. The chirality of helix chain
may be characterized by the eigenvalues of $\hat{O}_{ijk}$,
$\hat{O}_{ijk}|\psi\rangle=\pm1|\psi\rangle$.

The Hamiltonian commutes with two elementary loop variables. For
loop A in Fig. 1(d), the conserved loop variable $P_A$ is given by
\begin{eqnarray}
\hat{P}_{A}=\sigma^{x}_{7}\sigma^{y}_{8}\sigma^{z}_{9}\sigma^{x}_{11}
\sigma^{y}_{12}\sigma^{z}_{13}.
\end{eqnarray}
And the conserved loop variable $P_B$ in loop B (Fig 1(a)) is
given by
\begin{eqnarray}
 \label{B}
\hat{P}_{B}=\sigma^{y}_{1}\sigma^{z}_{2}\sigma^{z}_{3}\sigma^{z}_{4}
\sigma^{x}_{5}\sigma^{y}_{6}
\sigma^{z}_{7}\sigma^{z}_{8}\sigma^{z}_{9}\sigma^{x}_{10},
\end{eqnarray}
which is equivalent to the product of two hexagon plaquettes with
their sharing bonds popping out as inter-layer bonding (Fig. 1
(a)). The eigen values of them are $P_A=\pm 1$ and $P_B=\pm 1$
because $\hat P^2_A=\hat P_B^2=1$. Notice that $\hat P_A$ and
$\hat P_B$ may be simultaneously diagonalized because of $[\hat
P_A,\hat P_B]=0$.

According to these conserved loop variables, we may obtain the
exact ground state of this model by using Majorana fermions to
express spins \cite{kitaev},
\begin{equation}
\sigma^{x}_{\textbf{i}}=ib_{\textbf{i}}^{x}
c_{\textbf{i}},\;\;\sigma^{y}_{\textbf{i}}=ib^{y}_{\textbf{i}}c_{\textbf{i}},
\;\;\sigma^{z}_{\textbf{i}}=ib_{\textbf{i}}^{z}c_{\textbf{i}}
\end{equation}
with constraint $b_{x}b_{y}b_{z}c=1$. Majorana fermions
$b_{i}^{x},b_{i}^{y},b_{i}^{z},c_{i}$ obey the anti-commutations
relation
\begin{equation}
\{b_i^{\alpha},b_j^{\beta}\}=2\delta_{\alpha\beta}\delta_{ij}
,\;\;c_{i}^{2}=1,\;\;\{b_i^{\alpha},c_j\}=0.
\end{equation}
 The nearest
neighbor spin coupling term reads
$\sigma_{i}^{\alpha}\sigma_{j}^{\alpha}=b_{i}^{\alpha}b_{j}^{\alpha}c_{i}c_{j}$,
One may verify that the operator
$u^{\alpha}_{ij}=ib_i^{\alpha}b_j^{\alpha}$ commutes with the
Hamiltonian and other $u^{\beta}_{ij}$ operators, i.e.,
\begin{equation}
[u^{\alpha}_{ij},H_{d}]=0,\;\;[u^{\alpha}_{ij},u^{\beta}_{kl}]=0.
\end{equation}
Therefore, $u^{\alpha}_{ij}$ may be viewed as the bond operator
connecting two nearest neighboring lattice sites and they may be
though as 'good quantum numbers' which reflect the real physical
conserved loop variables $P_A$ and $P_B$. To see this point, we
define the Wilson loop by a series of product of $u_{ij}$ along a
closed path. There are two classes of elementary loops that can
not be decomposed into smaller ones in this model, which are
exactly the loop A and loop B in Fig. 1. The products of $u_{ij}$
along these loops are no more than the expressions of $\hat P_A$
and $\hat P_B$ in Majorana ferioms. Following Lieb's
theorem\cite{lieb}, the ground state must be vortex free state.
The eigenvalue of the hexagon plaquette operator $\hat{P}_{A}$ at
vortex free state is ${P}_{A}=1$. This plaquette operator may be
interpreted as the magnetic flux through the plaquette, so does
the double hexagon plaquette operator $\hat{P}_{B}$. The magnetic
field line is perpendicular to the minimal surface of the Wilson
loop. We may introduce magnetic helicity to characterize the
quantum entanglement between loops of different size.

The Wilson loop operators are good quantum numbers. They share the
same Hilbert space with Hamiltonian. Small loops do not gain extra
energy when they expand into larger configuration. So these loops
have no stress tension. They have the same ground state energy.
Any spin configuration along a closed path which preserves
$W_{c}=-1$ for Wilson loop defines an elementary excitation. The
excited state space is also highly degenerate. There are two types
of fundamental quasiparticles in honeycomb helix lattice model,
$P_{A}=-1$ and $P_{B}=-1$. The honeycomb helix lattice model has
nontrivial loop quasiparticles which obey exotic statistics in
three dimensions. Two loops confined in two dimensions could not
entangle each other without cutting them up. But in three
dimensions, they can avoid crossing each other when they wind
around each other by keeping their topology. If they are linked
with each other, the continuous transformation between different
loop configurations corresponds to the transition between
different eigenstates. Quantum entanglement in the degenerate
Hilbert space is the origin of non-abelian statistics.

The the vortex free ground state is given by $u_{ij}=+1$. The
Hamiltonian becomes quadratic
$H_{h}=\sum{h}_{i\mu,j\nu}c_{i,\mu}c_{j,\nu},$ where $i,j$ are the
site index of unit cell, $\mu,\nu$ indicates the internal degrees
inside the unit cell. In the honeycomb helix model, the unit cell
is a bounded pair of plaquette $A$ and plaquette $B$(Fig. 1). The
fundamental translational invariant vector of the unite cell is
$\textbf{r}_{i}=m\textbf{e}_{a}+n\textbf{e}_{b}+l\textbf{e}_{c}$
with basis vector $\textbf{e}_{a}=(3\sqrt{3}a,0,0)$,
$\textbf{e}_{b}=(\frac{3\sqrt{3}}{2}a,\frac{3}{2}a,\frac{1}{3}c)$
and $\textbf{e}_{c}=(0,0,c)$(Fig. 1). Performing the Fourier
transformations,
\begin{equation}
h_{\mu\nu}(\textbf{q})=\sum_{j}{e}^{i(\textbf{q},\textbf{r}_{j})}h_{0\mu,j\nu},\;\;c_{\textbf{q},\mu}=\frac{1}{\sqrt{N}}\sum_{j}e^{-i(\textbf{q},\textbf{r}_{j})}c_{j\mu},
\end{equation}
we can derive the momentum representation of the Hamiltonian. The
exact spectrum is
\begin{eqnarray}\label{spectrum}
&&E_{g}(J_{x},J_{y},J_{z},\textbf{{q}})=(J_{x}^{6}-2\cos{3
\sqrt{3} a q_{x}} J_{x}^{3} J_{y}^{3}+J_{y}^{6}\nonumber\\
&&+2\cos({\frac{3}{2} \sqrt{3}} a q_{x}+\frac{3 a
q_{y}}{2}-\frac{2}{3} c q_{z})J_x^4 {J}_{y} J_z \nonumber\\
&&+4 \cos(\frac{3}{2}\sqrt{3}a q_x + \frac{3}{2}a q_y +
\frac{c}{3} q_z) J^4_x {J}_{y} J_z\nonumber\\
&&-4 \cos(\frac{3}{2} \sqrt{3} a q_x-\frac{3 a q_y}{2}-\frac{c
q_z}{3})J_x {J}_y^4 J_z \nonumber\\
&&-2\cos(\frac{3}{2} \sqrt{3} a q_x -\frac{3 a q_y}{2} + \frac{2 c
q_z}{3}) J_{x} J_y^4 J_z + 5 J_x^2 J_y^2 J_z^2 \nonumber\\
&&+4 \cos(c q_z) J_x^2 J_y^2 J_z^2 +2\cos(\frac{3}{2} \sqrt{3} a
q_x-\frac{9 a q_y}{2})J_x^3 J_z^3\nonumber\\
&& -2\cos(\frac{3}{2} \sqrt{3} a q_x - \frac{9 a q_y}{2})J_y^3
J_z^3 \nonumber\\
&&+4\cos(3 a q_y - (c q_z)/3)
 J_{x} J_{y} J_z^4 \nonumber\\
&&+2\cos(3 a q_y +\frac{2 c q_z}{3}) J_x J_y J_z^4 +
J_z^6)^{\frac{1}{6}}. \label{11}
\end{eqnarray}

 \begin{figure}
\begin{center}\label{phase}
\includegraphics[width=0.30\textwidth]{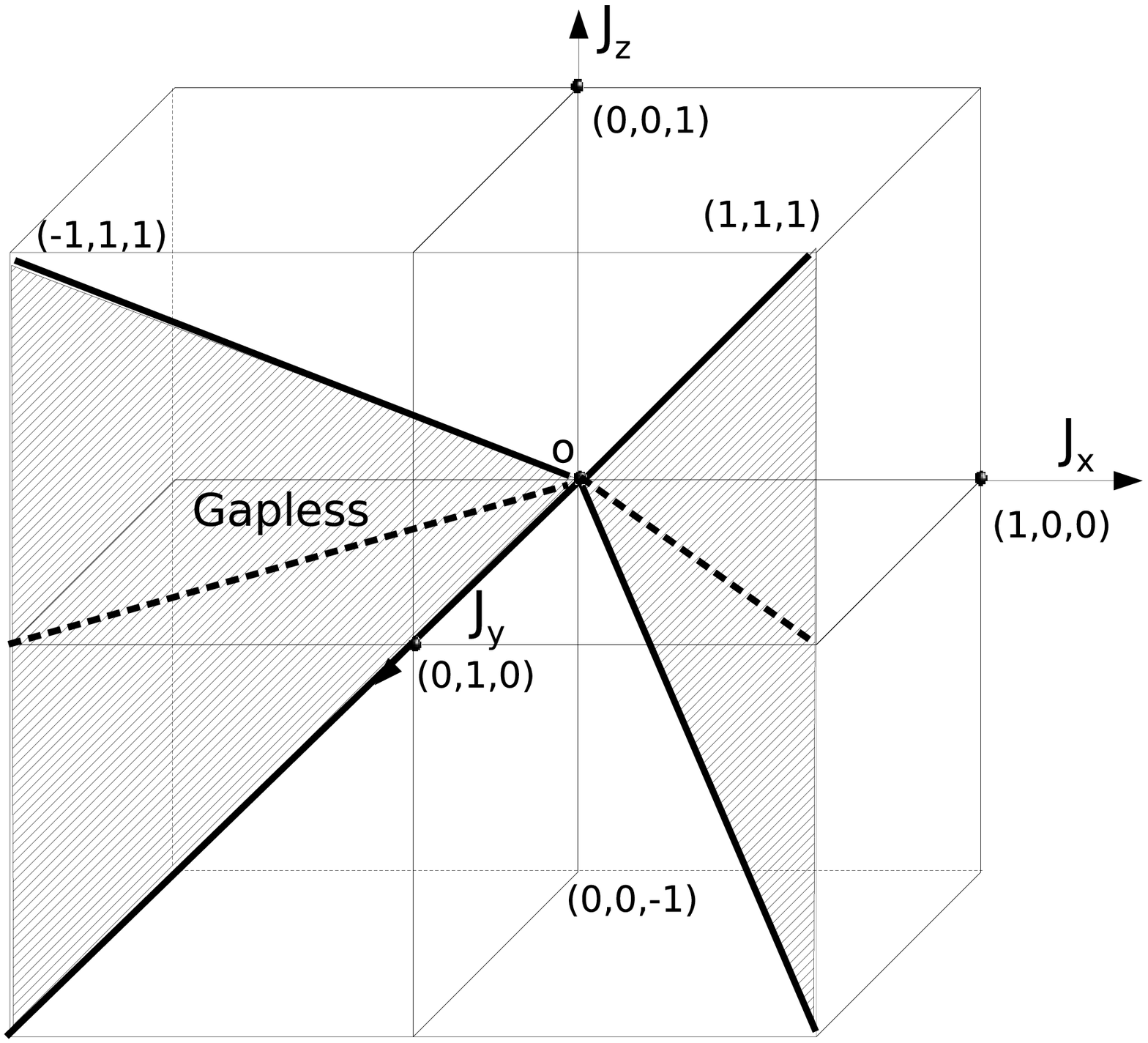}
\caption{The phase diagram of the honeycomb helix lattice model.
The shaded region is the gapless phase.}
\end{center}
\end{figure}

\begin{figure}
\begin{center}\label{xy}
\includegraphics[width=0.45\textwidth]{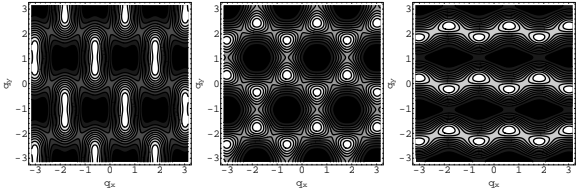}
\caption{A cross section of the energy spectrum in $q_{x}-q_{y}$
plane. The left panel is in gapless  phase, the fermions fill in
the low energy region while high energy region is cycled by low
energy states. This means there exists closed Fermi surface. The
middle panel is the critical region, the Fermi surface is nearly
opened. The right panel is in gapped phase. The Fermi surface is
completely opened. The dark area is the low energy region which is
filled by fermions. The bright area indicates the higher energy
region.}
\end{center}
\vspace{-1cm}
\end{figure}

The ground state is highly degenerate due to innumerable loop
operators which are integral of motion. The whole Hilbert space
are divided into two regimes---the gapped phase and gapless phase.
The gapless phase falls in the diagram covered by the
inequalities:
\begin{eqnarray}\label{honey-inequality}
|J_{i}|^{3}&\leq&3|J_{i}J_{j}J_{k}|+|J_{j}|^{3}+|J_{k}|^{3},\nonumber\\
3|J_{x}J_{y}J_{z}|&\leq&|J_{x}|^{3}+|J_{y}|^{3}+|J_{z}|^{3},
\end{eqnarray}
where $(i,j,k=x,y,z;\;i\neq{j}\neq{k})$. The gapless phase region
in the phase diagram is an infinite triangular area in the plane
of $|x|=|y|$(Fig. 2).

The quadratic Hamiltonian expressed by Majorana fermions operator
has an equivalent representation in terms of complex fermion
operators. We first express the spins by Majorana operators
$\sigma^{\alpha}_{\textbf{i}}=ib_{\textbf{i}}^{\alpha}
c_{\textbf{i}}$, and then introduce the complex fermion\cite{chen}
on the $\sigma^{z}_{i}\sigma^{z}_{j}$ coupling bonds,

\begin{equation}
d_{r}=(c_{i}+ic_{j})/2,\;\;d_{r}^\dag=(c_{i}-ic_{j})/2.
\end{equation}

The spin-spin coupling terms in Hamiltonian can be transformed
into standard fermion representation,
\begin{eqnarray}
&&\sigma^{z}_{i}\sigma^{z}_{j}=u^{z}_{ij}(2d_{\textbf{r}}^\dag{d}_{\textbf{r}}-1),\nonumber\\
&&\sigma^{x}_{i}\sigma^{x}_{j}=u^{x}_{ij}(d_{\textbf{r}}^{\dag}+d_{\textbf{r}})(d_{\textbf{r}-{\textbf{e}}_x}^{\dag}-d_{\textbf{r}-{\textbf{e}}_x})\nonumber\\
&&\sigma^{y}_{i}\sigma^{y}_{j}=u^{y}_{ij}(d_{\textbf{r}}^{\dag}+d_{\textbf{r}})(d_{\textbf{r}-{\textbf{e}}_y}^{\dag}-d_{\textbf{r}-{\textbf{e}}_y})
\end{eqnarray}
where $\textbf{r}$ denotes the center of
$\sigma^{z}_{i}\sigma^{z}_{j}$ bonds and
$u^{\alpha}_{ij}=ib_{i}^{\alpha}b_{j}^{\alpha}$. The Hamiltonian
(\ref{Hexagon-helix}) in the vortex free configuration is
transformed into quadratic Hamiltonian expressed by standard
fermions. The Fermi surface may be figured out from (\ref{11}) and
has different topological behavior in gapped phase and gapless
phase. We plot a cross section of the Fermi surface in Fig. 3. The
center of the dark disc area is the lowest energy point. The
bright area is the high energy region. The Fermi surface in the
left panel of Fig. (3)  is periodically arranged bubbles in three
dimensional momentum space. Their cross section at $q_{x}-q_{y}$
plane are closed circles without intersection. The Fermi surface
in the gapless phase are topologically equivalent to compact
sphere in momentum space. The middle panel is critical phase, at
which the closed Fermi surface began touching each other but not
connected. The right panel of Fig. 3 is in the gapped phase, the
closed Fermi surface of gapless phase opened and connected each
other(Fig. 3).

The most natural topological index to distinguish the two phases
is Hopf index which defines a map of from $S^{3}$ in momentum
space to a sphere $S^{2}$ in coherent function space. The
nontrivial two dimensional compact manifold embedded in the three
dimensional momentum space comes from the gapless solution
manifold $E(p)=0$. The Hopf index is an integer which counts how
many time the sphere in momentum space wraps the target sphere
following the map $f_{Hopf}:S^{3}(p)\rightarrow{S^{2}}$. The
nontrivial Hopf index means that the sphere enclose a monopole in
momentum space. In the gapless phase, monopoles are enclosed by
fermi sphere. They are at the right center of the dark disc areas
in Fig. 3. They have either positive charge or negative. The
monopole-anti-monopole pairs are fused into vacuum in the gapped
phase, so we can not detect them through the Hopf index. The
critical region occurs when the monopole-anti-monopole began
touching each other but is not near enough to annihilate.

The topological quantum phase transition from confined monopole
phase to deconfined monopole phase leads to the divergence of the
second order derivative of the ground state energy on the critical
boundary. The divergence behavior of approaching the critical
point from the gapped phase is totally different from that of
gapless phase(Fig. 4). In the gapped phase, a sharp peak far from
the critical point is observed in the region of $J_{z}>0$. This
peak is finite, it indicates a crossover transition.

\begin{figure}
\begin{center}\label{spiral}
\includegraphics[width=0.4\textwidth]{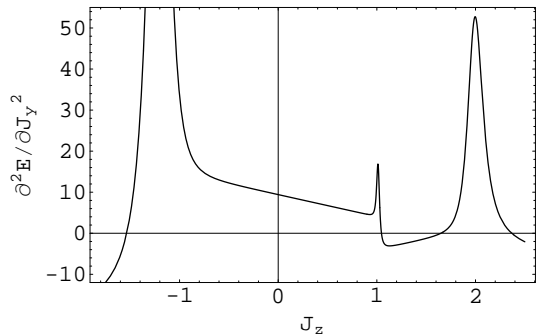}
\caption{The second order derivative of the ground state spectrum
at $J_{x}=J_{y}=1$. In the region $-1<J_{z}<1$ is gapless phase
and outside is gapped phase.}
\end{center}
\end{figure}

Now we turn to the spin correlations in the honeycomb helix
lattice model. As shown in Ref. \cite{baskaran} on Kitaev model of
two dimensional honeycomb lattice, the two-spin correlation
vanishes beyond nearest neighbor separation in different bonding
directions. In the three dimensional honeycomb helix lattice
model, the two-spin correlation demonstrated the same behavior.
The representation of spins by Majorana fermions factorized the
eigenstates of physical Hilbert space as gauge sector and
corresponding matter sector,
$|\psi\rangle=|\psi_{Mg}\rangle|\psi_{g}\rangle$ with
\begin{equation}
\hat{u}_{ij}|\psi_{g}\rangle\equiv{b_{\textbf{i}}^{\alpha}
b_{\textbf{j}}^{\alpha}}|\psi_{g}\rangle=u|\psi_{g}\rangle.
\end{equation}
The spin operator $\sigma^{\alpha}_{i}$ acting on any eigenstate
adds a Majorana fermion at site $i$ and one $\pi$ flux each to the
plaquettes intersecting with the bond
$\langle{ij}\rangle^{\alpha}$. The two-spin correlation function
\begin{equation}
C^{\alpha\beta}_{\textbf{i}\textbf{j}}(t)=\langle\psi_{g}|\langle\psi_{Mg}|\sigma^{\alpha}_{\textbf{i}}(t)\sigma^{\beta}_{\textbf{j}}|\psi_{Mg}\rangle|\psi_{g}\rangle
\end{equation}
measures the probability we will detect the added Majorana fermion
and flux at lattice site $j$. If Hilbert spaces of the two-spins
at ($\textbf{i},\textbf{j}$) have no overlap on the gauge sector,
the two-spin correlation is constantly zero. In this three
dimensional honeycomb helix model, the two spins at
($\textbf{i},\textbf{j}$) have no overlap in Hilbert space unless
they are nearest neighbors. Therefore two-spin correlation
vanishes beyond nearest neighboring separation.

The correlation between plaquette excitations is actually
multispin correlation. The non-vanishing multispin correlation
comes from the operators which are product of an arbitrary number
of the terms that appear in the Hamiltonian. We can compute the
multispin correlation using the method in Ref. \cite{baskaran}.
Here we first consider the correlation between the two types of
elementary plaquette excitations. As shown in Fig. 1, a spin
starts from white lattice site $\textbf{i}$ has three different
propagating direction. The interlayer spin correlation
$C^{\alpha\beta}_{\textbf{i}\textbf{i}'}(t)$ are sharply cut off
at nearest neighbor separation. This two-spin correlation measures
the coupling between two $\hat{P}_{b}$ plaquette excitations. The
coupling between $\hat{P}_{a}$ plaquette excitation and
$\hat{P}_{b}$ plaquette excitation is measured by three-spin
correlation. One spin flip at $\textbf{i}$ excites two plaquette
excitations, $\hat{P}_{A}=-1$ and $\hat{P}_{b}=-1$. Their sharing
boundary is along
$\textbf{i}\rightarrow\textbf{j}\rightarrow\textbf{k}$. For a spin
flip propagating along the path
$\textbf{i}\rightarrow\textbf{j}\rightarrow\textbf{k}$, the
nonzero three-spin correlation
\begin{equation}\label{3spins}
C^{\alpha\beta\gamma}_{\textbf{i}\textbf{j}\textbf{k}}(t)=\langle\psi_{g}|\langle\psi_{Mg}|
\sigma^{\alpha}_{\textbf{i}}(t)\sigma^{\beta}_{\textbf{j}}(t)
\sigma^{\gamma}_{\textbf{k}}(t)|\psi_{Mg}\rangle|\psi_{g}\rangle
\end{equation}
ends at another white site $\textbf{k}$. The two elementary
plaquette operators are the smallest loops, they can not be
decomposed into smaller ones. As for the more complex loop-loop
correlation, we shall introduce non-abelian Chern-Simons theory in
the following sections.

We introduce the other three dimensional helix models in the
following. They have more or less similar property as the
honeycomb helix model. We left the exact solution and present only
the spin-spin coupling configuration.

\subsection{The Square Helix model}
\begin{figure}[htbp]
\centering
\par
\begin{center}
$%
\begin{array}{c@{\hspace{0.01in}}c}
\includegraphics[scale=0.35]{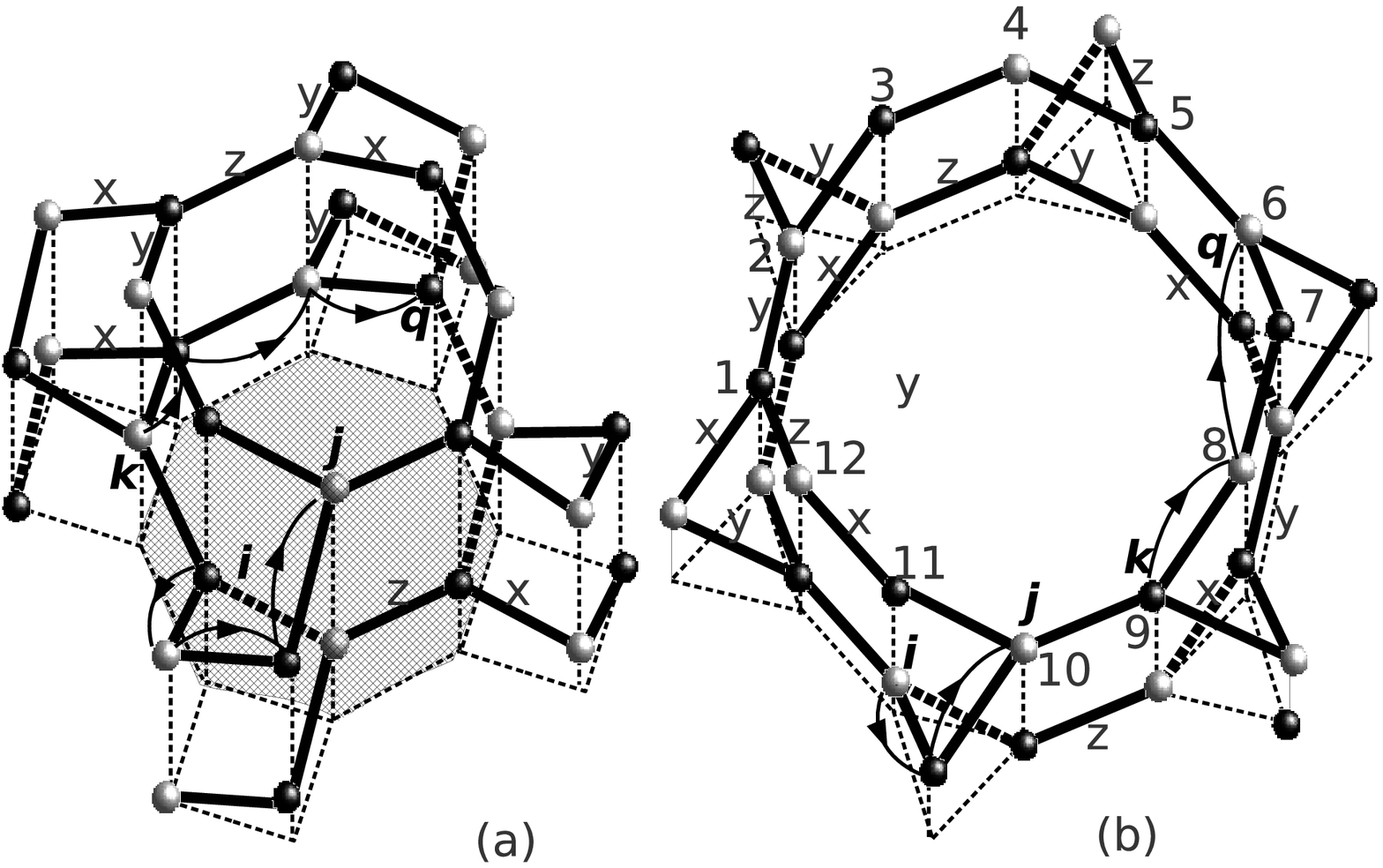}\\ %
\mbox{\bf (I)} & \mbox{} \\
\includegraphics[scale=0.35]{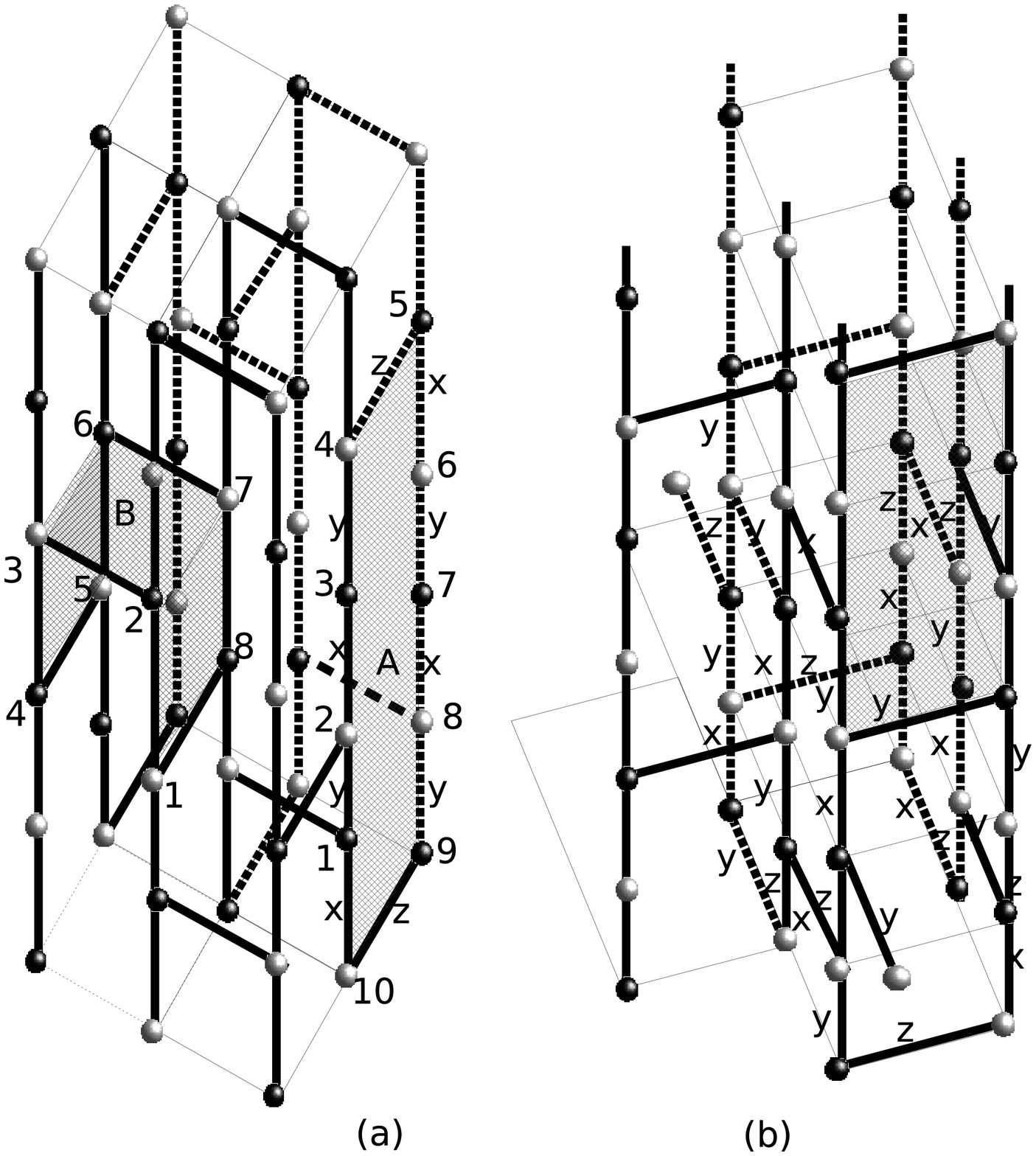} \\
\mbox{\bf (II)} & \mbox{}%
\end{array}%
$%
\end{center}
\caption{{\protect\small (I)-(a) The square helix lattice model.
(I)-(b) The triangle helix model on Cu-sublattice in
${C{u}}_{6}{S{i}}_{6}{O}_{3}\cdot6{H_{2}O}$.}} (II)-(a) An
equivalent demonstration of square helix model on cubic lattice.
(II)-(b) The deformed triangle helix model on cubic lattice.
\label{square22}
\end{figure}

The square helix model is composed of alternative $XY$ spin chains
each of which has four nearest neighboring chains coupled to one
another through $\sigma^{z}\sigma^{z}$ bonding(Fig. 5, (a)). The
Hamiltonian includes two parts, $H_{SH}={\sum}H_{xy}+H_{z}$,
\begin{eqnarray}\label{square-helix}
H_{SH}&=&\sum_{{ij,k=even}}[J_{y}\sigma^{y}_{ij,k-1}\sigma^{y}_{ij,k}+J_{x}\sigma^{x}_{ij,k}\sigma^{x}_{ij,k+1}]\nonumber\\
&+&\sum_{i{\pm}n,j,k{\pm}2n}[J_{z}\sigma^{z}_{i-1,j,k}\sigma^{z}_{i,j,k}]\nonumber\\
&+&\sum_{i,j{\pm}n,k{\pm}2n}[J_{z}\sigma^{z}_{i,j,k+1}\sigma^{z}_{i,j+1,k+1}].
\end{eqnarray}
A single helix chain is exactly the same single $XY$ spin chain in
two dimensional Kitaev model, which is equivalent to one
dimensional Ising model with transverse field\cite{feng}. The
multispin correlation function along a single helix chain
demonstrates a long range order,
\begin{equation}\label{sxx}
S^{xx}_{2j}=\lim_{j\rightarrow\infty}\langle\prod^{2j}_{k=1}\sigma^{x}_{k}\rangle\sim[1-(J_{y}/J_{x})^{2}]^{1/4}.
\end{equation}
In the three dimensional square-helix model, it is the product
operator of the bonding terms in Hamiltonian that has nonzero
multispin correlation. While two-spin correlation in real lattice
space vanishes at the nearest neighboring
separation\cite{baskaran}. The chiral operator is the combination
of different product of three spin operators, it is a good
quantity to characterize the quantum entanglement of three spins.

The Hamiltonian (\ref{square-helix}) is exactly soluble following
a similar procedure as that in honeycomb helix model except a
different gauge fixing. There are two types of fundamental
plaquettes in square helix model. It is more clear to see the
picture by deforming the square helix lattice into cubic
lattice(Fig. 5). The energy minimum of ground state is achieved by
$0-\pi-0-\pi\cdots$ alternative flux pattern following Lieb's
theorem. The integrals of motion $P_{A}$ take a phase $2\pi$ for
all irreducible plaquette $\emph{A}$,
$\hat{P}_{A}=\sigma^{z}_{1}\sigma^{z}_{2}\sigma^{z}_{3}
\sigma^{x}_{4}\sigma^{y}_{5}\sigma^{z}_{6}
\sigma^{z}_{7}\sigma^{z}_{8}\sigma^{x}_{9}\sigma^{y}_{10}=e^{i2\pi}$,
and for all plaquette $\emph{B}$,
$\hat{P}_{B}=\sigma^{y}_{1}\sigma^{y}_{2}\sigma^{x}_{3}
\sigma^{x}_{4}\sigma^{y}_{5}\sigma^{y}_{6}\sigma^{x}_{7}
\sigma^{x}_{8}=e^{i\pi}$.

Elementary plaquette excitations are produced by flipping odd
number of spins along the plaquettes $\emph{A}$ and $\emph{B}$. At
the excited states, the phase pattern of the two palquettes are
$P_{A}=-1$ and $P_{B}=+1$. The highly degenerate eigenstates are
classified by different spin configurations. The spin flip at
lattice site $\textbf{i}$ excites four $\pi$ flux each in its four
neighboring plaquettes. Four plaquette excitations are always
generated and annihilated at the same time. So there are many
degree of freedom to operate on them. Two of the four plaquette
excitations may share a strong four-spin correlation in
$k$-direction,
\begin{equation}
C^{yzxx}_{k,k+1,k+2,k+3}=
\langle\sigma^{y}_{ij,k}\sigma^{z}_{ij,k+1}\sigma^{x}_{ij,k+2}\sigma^{x}_{ij,k+3}\rangle.
\end{equation}
This four-spin correlation only covers four neighboring lattice
sites,
\begin{equation}
C^{yzx\lambda}_{k,k+1,k+2,k+m}(t)=0, m>3.
\end{equation}
While another two of the four plaquette excitation are connected
only by nonzero two-spin correlation. Thus the correlation between
plaquette excitations is different along different direction.

\subsection{The triangle helix model on Dioptast Lattice}

The triangle helix model is established on a three dimensional
lattice which exist in nature. The $Cu$ atoms in dioptase
${C{u}}_{6}{S{i}}_{6}{O}_{3}\cdot6{H_{2}O}$ construct the exact
triangle helix lattice, $Cu$-atoms form chiral chains along
$z$-axes and project a honeycomb configuration to the $X-Y$
plane\cite{gros}.

The exact solution of the Heisenberg spin-spin coupling
hamiltonian on the dioptase lattice is hardly reachable. We
derived an exactly soluble model by introducing Kitaev-type
spin-spin coupling interactions along different bonds on this
three dimensional lattice(Fig. 2(b)). The projection of the
triangle helix model to $X-Y$ plane is a two dimensional chiral
spin liquid model constructed in Refs. \cite{yao}\cite{yangsun}
which has a stable ground state spontaneously breaking time
reversal symmetry. While our three dimensional triangle helix
model is invariant under time reversal transformation. The two
elementary plaquettes, octagon $\hat{P}_{8}$ and dodecagon
$\hat{P}_{12}=\sigma^{x}_{1}\sigma^{z}_{2}\sigma^{y}_{3}
\sigma^{x}_{4}\sigma^{z}_{5}\sigma^{y}_{6}
\sigma^{x}_{7}\sigma^{z}_{8}\sigma^{y}_{9}
\sigma^{x}_{10}\sigma^{z}_{11}\sigma^{y}_{12}$, are both in
$\pi-$flux phase at ground states. Unlike the quantum phase
transition from anti-ferro-magnetically ordered state to a quantum
spin liquid in magnetic crystal
${C{u}}_{6}{S{i}}_{6}{O}_{3}\cdot6{H_{2}O}$, the quantum phase
transition here in the triangle helix model is between two
topologically distinguished spin liquid states.

\subsection{The Square-Hexagon-Helix Lattice Model}

The square-hexagon-helix lattice model projects one of the two
dimensional models\cite{yangsun} with Kitaev type's coupling in
$X-Y$ plane. The hexagon helix with opposite chirality are
separated by isolated squares. The spatial distribution of the
helixes is a pillar array on Kagome lattice(Fig. 6). Further
observation shows, the square-hexagon-helix model is composed of
three coupled Ising models which are established on multilayer
Kagome lattice.

\begin{figure}[htbp]
\centering
\par
\begin{center}
$%
\begin{array}{c@{\hspace{0.01in}}c}
\multicolumn{1}{l}{\mbox{}} & \multicolumn{1}{l}{\mbox{}} \\
\includegraphics[scale=0.25]{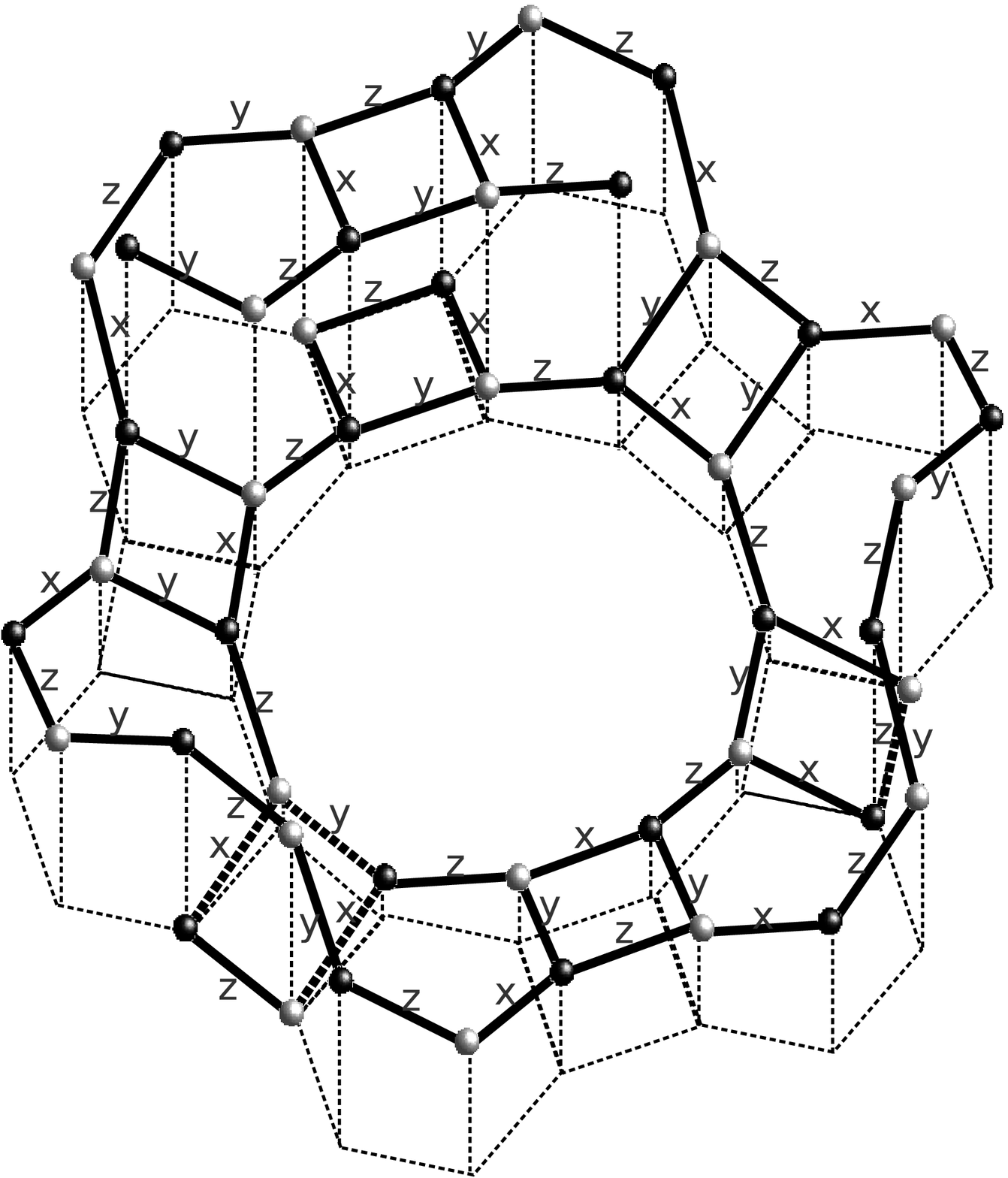} & %
\includegraphics[scale=0.2]{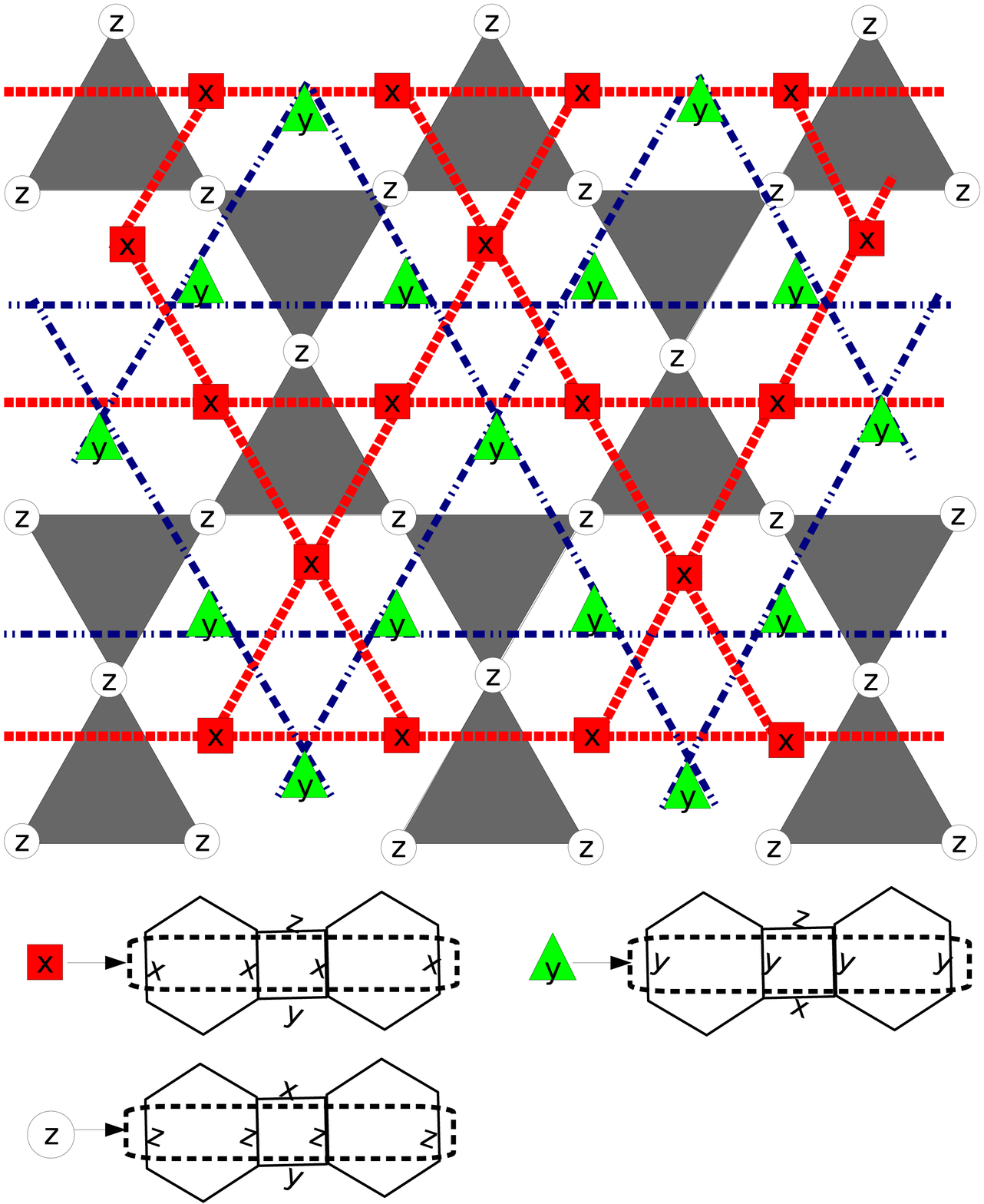} \\
\mbox{\bf (a)} & \mbox{\bf (b)}%
\end{array}%
$%
\end{center}
\caption{(a) The square-hexagon-helix Lattice model. (b) An
equivalent configuration of square-hexagon-helix Lattice model on
multi-layers Kagome lattice.} \label{tricomb}
\end{figure}

\section{Three dimensional chiral spin liquid models}

In the sections above, we have derived four typical spin models
which are exactly soluble in three dimensions. Their ground states
with short range interaction are spin liquid states. It has been a
long history to look for three dimensional chiral spin liquid, so
far there is still no exact soluble three dimensional model to
confirm its existence. In this section, we shall construct a
series of three dimensional chiral spin liquid models. They are
exactly solvable. Their exact stable ground states spontaneously
break the time reversal symmetry. These chiral spin liquids are
likely to be realized by ultracold atoms trapped in cubic optical
lattice.

\subsection{Chiral Square Helix Model}

We add two spins at all of the horizontal $\sigma^{z}\sigma^{z}$
bonds in the cubic version of deformed square helix lattice(Fig. 5
(II)-(a)). One is at $1/3$ of one lattice spacing and the other is
at $2/3$. We shift the added spins vertically a half lattice
spacing along $z$-axis. If the spins are assumed to have only the
nearest neighboring interaction, we get the first chiral spin
liquid in three dimensions(Fig. 7 (a)). The Hamiltonian of the
chiral square helix reads

\begin{eqnarray}\label{chiral-cubic-Hamilton}
H_{cc}&=&-\sum_{{ij,k=even}}[J_{y}\sigma^{y}_{ij,k-\frac{3}{4}}\sigma^{y}_{ij,k-\frac{1}{4}}+J_{z}\sigma^{z}_{ij,k-\frac{1}{4}}\sigma^{z}_{ij,k+\frac{1}{4}}\nonumber\\
&+&J_{x}\sigma^{x}_{ij,k+\frac{1}{4}}\sigma^{x}_{ij,k+\frac{3}{4}}
+J_{z}\sigma^{z}_{ij,k+1-\frac{1}{4}}\sigma^{z}_{ij,k+1+\frac{1}{4}}]\nonumber\\
&+&\sum_{i{\pm}n,j,k{\pm}2n}[J_{y}\sigma^{y}_{i-\frac{1}{3},j,k}\sigma^{y}_{ij,k+\frac{1}{4}}
+J_{x}\sigma^{x}_{i-\frac{1}{3},j,k}\sigma^{x}_{ij,k-\frac{1}{4}}\nonumber\\
&+&J_{z}\sigma^{z}_{i-\frac{2}{3},j,k}\sigma^{z}_{i-\frac{1}{3},j,k}\nonumber\\
&+&J_{y}\sigma^{y}_{i-1,j,k+\frac{1}{4}}\sigma^{y}_{i-\frac{2}{3},j,k}
+J_{x}\sigma^{x}_{i-1,j,k-\frac{1}{4}}\sigma^{x}_{i-\frac{2}{3},j,k}]\nonumber\\
&+&\sum_{i,j{\pm}n,k{\pm}2n}[J_{y}\sigma^{y}_{i,j,k+\frac{3}{4}}\sigma^{y}_{i,j+\frac{1}{3},k+1}\nonumber\\
&+&J_{x}\sigma^{x}_{i,j,k+1+\frac{1}{4}}\sigma^{x}_{i,j+\frac{1}{3},k+1}\nonumber\\
&+&J_{z}\sigma^{z}_{i,j+\frac{1}{3},k+1}\sigma^{z}_{i,j+\frac{2}{3},k+1}\nonumber\\
&+&J_{y}\sigma^{y}_{i,j+\frac{2}{3},k+1}\sigma^{y}_{i,j+1,k+\frac{3}{4}}
+J_{x}\sigma^{x}_{i,j+\frac{2}{3},k+1}\sigma^{x}_{i,j+1,k+\frac{5}{4}}].\nonumber\\
\end{eqnarray}
Our interest focus mainly on the gauge symmetry of exact ground
state. As shown in Fig. 7 (a), there are three elementary
plaquettes: a triangle and two polygons. One of the two polygons
has sixteen bonds, the other covers twenty bonds. The energy is
minimized by uniform flux configuration. The elementary plaquettes
are all in $\pi$-flux phase at ground state. The $\pi-$flux of
triangular plaquette flips under time reversal transformation.
Although the two polygon plaquettes are invariant under time
reversal transformation, the global flux pattern spontaneously
breaks the time reversal symmetry. Thus the ground state is highly
degenerate chiral spin liquid.

The fractionalized elementary excitations are loop excitations in
the three dimensional chiral spin liquid. Large loop operator is
the product of the elementary plaquettes covered by its minimal
surface. If the loop has odd number of links, there must be odd
number of triangular plaquettes confined in the area surrounded by
the loop. A spin flip adds $\pi$-flux each to the plaquettes
intersecting at that point. To some extend, the time reversal
transformation behaves like transition operator between excited
states and ground states. The loop with odd number of particles
spontaneously breaks the time reversal symmetry.

The correlation between plaquette excitations is measured by
multispin correlation which are spatially anisotropic. For
example, a spin flip at site $\textbf{q}$ excites four plaquettes
vertices sharing a bond along $z$-axis. Two of them share a strong
six-spin correlation
\begin{equation}
C_{\textbf{q}\textbf{q}_{1}\textbf{q}_{2}\textbf{q}_{3}\textbf{q}_{4}\textbf{p}}^{\alpha\beta\gamma\lambda\eta\kappa}
=\langle\sigma^{\alpha}_{\textbf{q}}\sigma^{\beta}_{\textbf{q}_{1}}\sigma^{\gamma}_{\textbf{q}_{2}}
\sigma^{\lambda}_{\textbf{q}_{3}}\sigma^{\eta}_{\textbf{q}_{4}}\sigma^{\kappa}_{\textbf{p}}\rangle.
\end{equation}
While another pair of plaquette excitations are only connected by
two-spin correlation which are sharply cut off beyond nearest
neighbors. This suggest that the fusion rules for plaquette
excitations in this mode might be more complex than that in two
dimensions.

\begin{figure}[htbp]
\centering
\par
\begin{center}
$%
\begin{array}{c@{\hspace{0.01in}}c}
\multicolumn{1}{l}{\mbox{}} & \multicolumn{1}{l}{\mbox{}} \\
\includegraphics[scale=0.25]{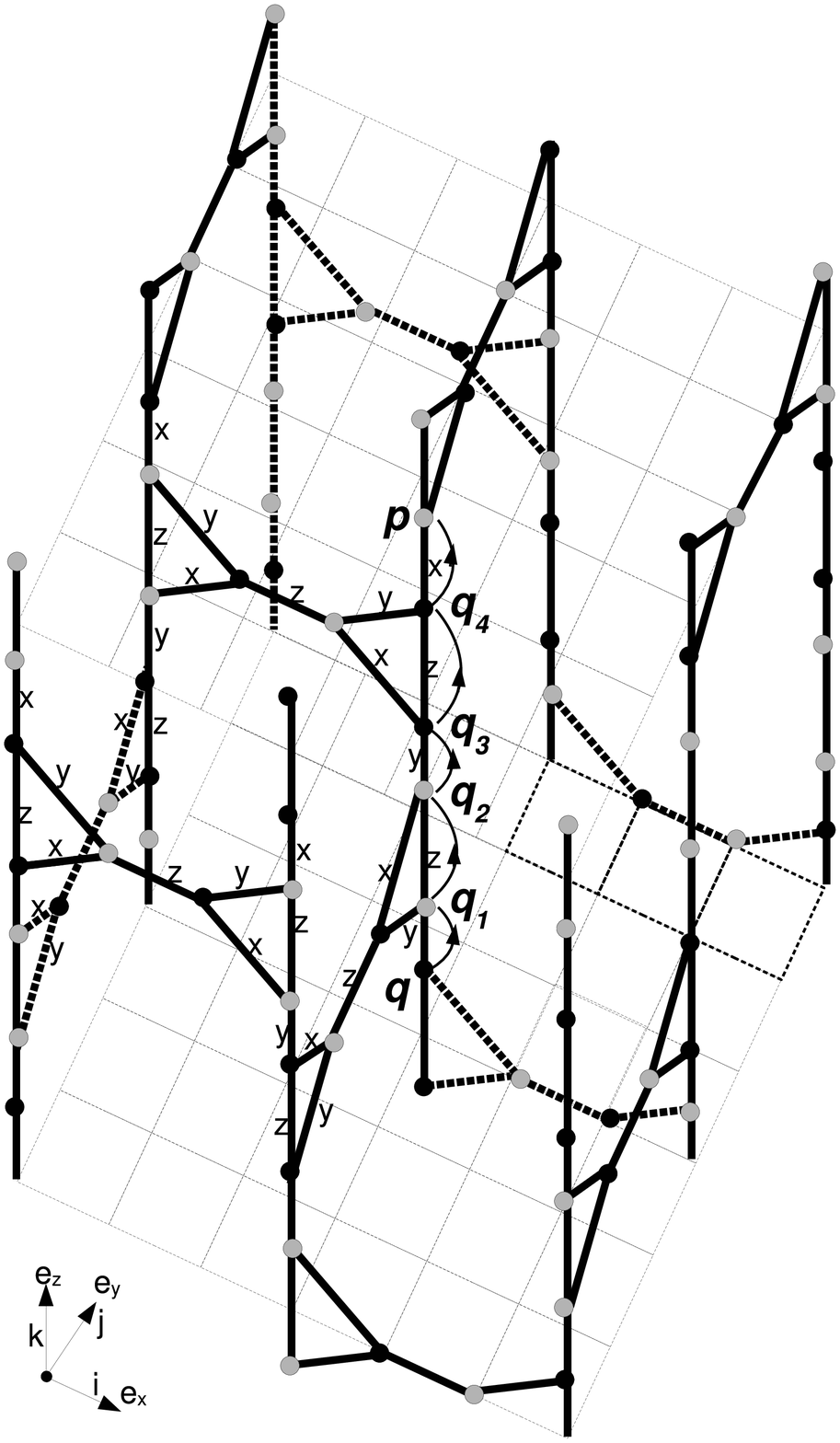} & %
\includegraphics[scale=0.25]{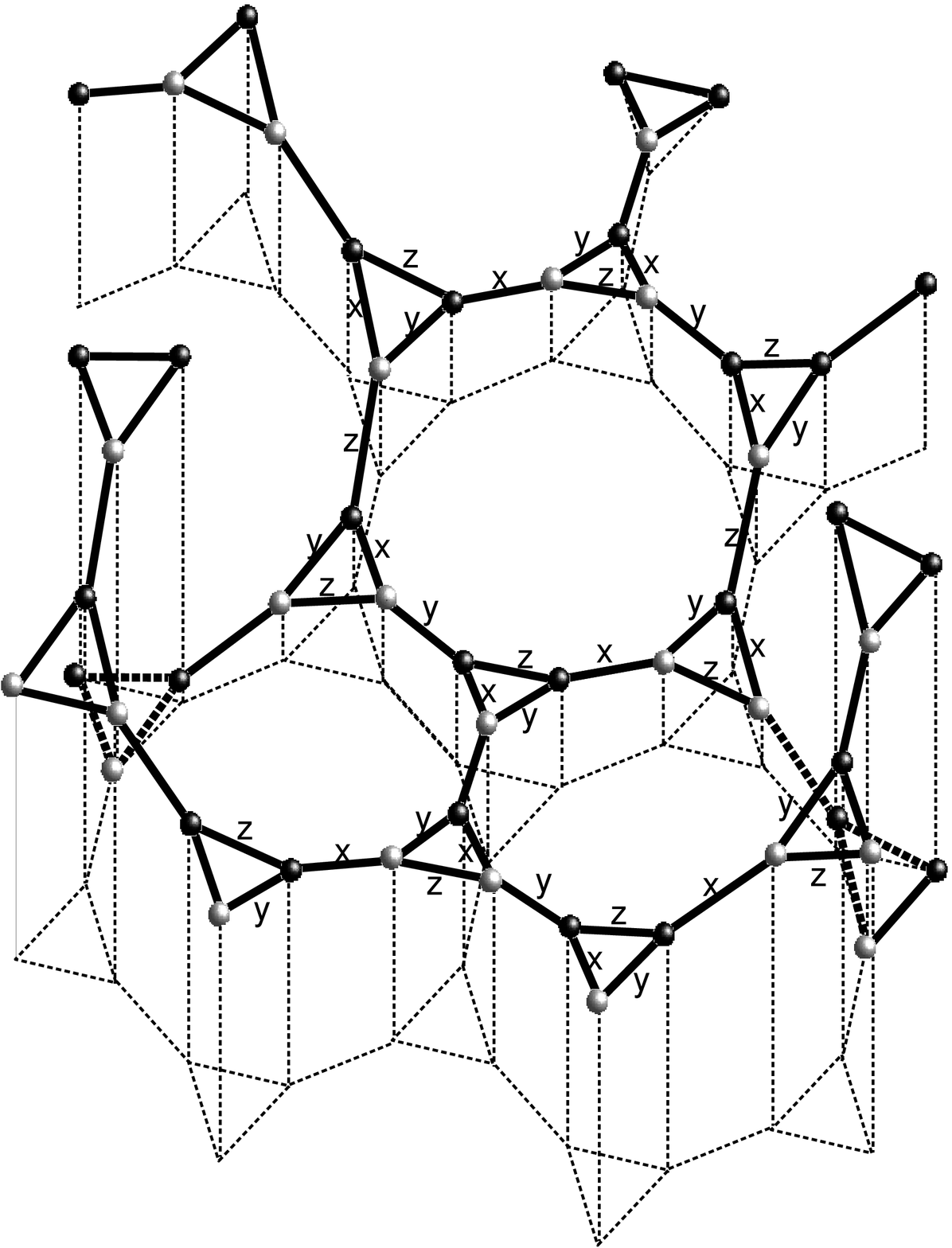} \\
\mbox{\bf (a)} & \mbox{\bf (b)}%
\end{array}%
$%
\end{center}
\caption{{\protect\small (a) The chiral square helix model on
cubic lattice. (b) The chiral dodecagon helix model. The character
structure is the dodecagonal helix.}} \label{chiiralspin}
\end{figure}

\begin{figure}
\begin{center}\label{square4}
\includegraphics[width=0.5\textwidth]{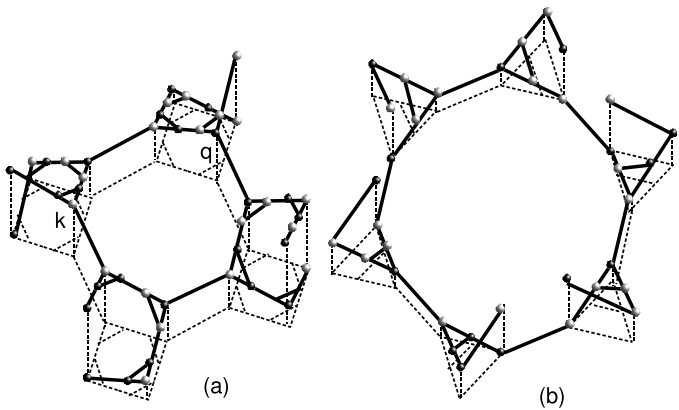}
\caption{(a) The chiral square helix model on the original square
helix lattice. (b) The chiral spin liquid model on the triangular
decorated dioptast lattice.}
\end{center}
\end{figure}

\subsection{The general principal to construct other artificial chiral spin liquid model}

There are many other ways to construct three dimensional chiral
spin models with Kitaev's type of coupling, which requires that
each lattice site branches out only three different bonds
corresponding to $\sigma_{x}\sigma_{x}$, $\sigma_{y}\sigma_{y}$
and $\sigma_{z}\sigma_{z}$. For example, we add two spins at every
bond to divide it into three equal sections, and assume spins only
interact with nearest neighbors, it naturally leads to a chiral
spin liquid in three dimension. In Fig. 8, we present two
examples, the triangular decorated square helix model and triangle
helix model. If one deforms the triangular decorated square helix
lattice model(Fig. 8 (a)) into cubic lattice, the triangular
decorated square helix is exactly the chiral square helix model.
One can also construct other models by dividing some selected
bonds into $n(n=2,3,4...)$ identical pieces and adding spins on
the adjoint points with properly arranged bonding direction. The
triangular decorated triangle-helix model(Fig. 8 (b)) is one of
such constructions, to construct this model, we add one spin each
on the two bonds branched from the same lattice site.

Another different way of constructing is to spiral polygon
plaquettes up. For example, we construct the dodecagon helix
lattice model which has the same two dimensional projection to
$X-Y$ plane as the diopaste lattice. Here it is the dodecagon that
spirals up with triangles as bridge between dodecagonal
helixes(Fig. 7). We can also obtain this chiral spin liquid model
by replacing each lattice site in the honeycomb helix model(Fig.
1) with a triangle. If we keep replacing the lattice sites with
triangles time and time again, it finally leads to the fractal
lattice model. This fractal model is different from Bethe lattice.
There is no closed loop in Bethe lattice, but in this fractal
lattice model, it is all sorts of loops that construct the whole
lattice. These loops form the complete set of integrals of motion.
Scale invariance lays in the heart of these exactly soluble
fractal lattice models.

Now we see that three dimensional chiral spin liquid emerges
following a rather simple doping construction. The only thing we
need is local interaction between nearest neighboring spins. When
it comes to existing material in nature, it is possible get chiral
spin liquid by doping spins in conventional three dimensional
lattice.

\section{Topological Correlation of entangled loops in three dimensional chiral spin liquid}

In the exactly soluble spin models obtained above, a closed path
$C$ consists of a series of bonds. Defining
$\hat{h}^{\gamma}_{ij}=\sigma^{\gamma}_{i}\sigma^{\gamma}_{j}$ for
a given bond $\gamma$, the product of $\hat{h}^\gamma_{ij}$ along
$C$ is a $Z_2$ gauge invariant quantity which is nothing but the
Wilson loop $W_{C}=\prod_{\gamma\in C}\hat{h}^\gamma_{ij}$. The
Hilbert space is divided into many gauge equivalent class
$\{|\psi_{C}\rangle\}$ corresponding to $W_{C}$ with
\begin{equation}\label{Wc}
W_{C}|\psi_{C}\rangle=\pm|\psi_{C}\rangle.
\end{equation}
Quantum information is encoded in this Hilbert space. The gauge
invariant stats can only be distinguished by global operation,
which protects the quantum information from decoherence due to
local quantum entanglement with environment.

Non-abelian excitations in three dimensional chiral spin liquid
serve as elementary qubits to perform quantum computation. Basic
logic gate may be constructed by designed braiding operation. To
perform topological quantum computation, we shall operate on
excitations obeying non-abelian braiding statistics. Point
quasiparticles in three dimensions are either fermion or boson
and thus they are not good candidates. It is the loop excitations
that satisfy non-commutable relation during the unitary evolution
in ground state subspace.

We first consider the unknotted loops which are not entangled or
linked with each other. If the loop correlation is extended to
strong and weak coupling limit of lattice gauge theory, it shows
that the loop correlations
$S_{\alpha\beta}:=\langle{W_{C_{\alpha}}W_{C_{\beta}}}\rangle$
decay exponentially with distance. At high temperature, the decay
width is governed by the area $A$ encapsulated by the two loops,
$S_{\alpha\beta}\propto{e^{cA}}$. At low temperature, the total
perimeter $|C_{\alpha}|+|C_{\beta}|$ determines the decay
behavior,
$S_{\alpha\beta}\propto{e^{c(|C_{\alpha}|+|C_{\beta}|)}}$.

Loop-loop correlation is actually bipartite multi-spin correlation
which can be calculated using the method developed in Ref.
\cite{baskaran}. The area $\hat{A}_{\alpha}$ enclosed by loop
$C_{\alpha}$ is composed of a collection of elementary plaquettes.
The Wilson loop is just the product of all the elementary
plaquette operators
\begin{equation}
W_{C_{\alpha}}=\prod_{i}{\hat{P}_{i}^{a}},
\end{equation}
$a$ indicates the type of elementary plaquette. For example,
$a=1,2,3$ in the chiral square helix model, $C_{1}$ and $C_{2}$
represent the two polygons and $C_{3}$ represents the triangular
plaquette. The degenerate Hilbert space of loop is the direct
product of the subspace of all the elementary plaquettes in
$\hat{A}_{\alpha}$. For a given loop $C_{\alpha}$, there exist
many different $\hat{A}_{\alpha}$ such that
$\partial\hat{A}_{\alpha}=C_{\alpha}$. The dominate contribution
to the correlation of two loops comes from the minimal surface $M$
encapsulated between them. The boundary of the minimal surface is
just the two loops, $\partial{M}=C_{\alpha}+C_{\beta}$. We denote
its Hilbert space as $\Psi_{M}$, the two loop correlation reads
\begin{equation}\label{wab}
S_{\alpha\beta}(t)=\langle\Psi_{\beta}|e^{i(H-E)t}
\hat{W}_{C_{\alpha}}(t)\hat{W}_{C_{\beta}}\rangle|\Psi_{\alpha}\rangle,
\end{equation}
which has an explicit expression
\begin{equation}
S_{\alpha\beta}(t)=\langle\Psi_{g}|e^{i(H-E)t}\prod_{\langle{ij}\rangle_{\alpha}}\sigma^{\gamma}_{{i}}\sigma^{\gamma}_{{j}}
\prod_{\langle{kl}\rangle_{\beta}}\sigma^{\eta}_{{k}}\sigma^{\eta}_{{l}}|\Psi_{g}\rangle.
\end{equation}
Expressing the spins in terms of Majorana fermions, we may divide
the Hilbert space into different gauge sectors according to the
static $Z_{2}$ gauge field operators $u_{ij}\equiv
i{b_{\textbf{i}}^{\gamma}b_{\textbf{j}}^{\gamma}}$. The multispin
correlation can be computed on the gauge sectors\cite{baskaran}.
However this computation is hard to tell the physical picture of
topological entanglement between loops.

In order to see the topological correlation more clearly, we take
a more physical picture in the continuum limit. We view a loop as
an electric current source radiating magnetic field. The
topological correlation between loops is equivalent to the
topological entanglement between magnetic field circles. Obviously
no matter how far the loops are separated, the magnetic field
lines are always entangled. When it comes to the knotted loops
which are linked, entangled one anther, besides the conventional
loop correlation which decays as their spatial separation
increases, there exist topological correlation which relies on the
global topology of links instead of the average distance between
them. The topological correlation is immune to local fluctuation
in space. Each knotted Wilson loop operator projects a subspace,
this topological correlation is a good quantity to measure the
quantum entanglement between different sectors of the total
Hilbert space.

To study the topological correlation of linked, knotted loops, we
first take a preview in the continuum limit. Considering the Hopf
link, in which two unknotted rings locked each other up, there is
only one crossing point between loop $C_{\alpha}$ and the minimal
surface of loop $C_{\beta}=\partial{A}_{\beta}$. The two loops
have no intersection. For a more general case, one loop winds
around another many times, the winding number is identical to
intersecting points
 \begin{equation}
C_{\alpha}\cap{{A_{\beta}}}=\{P_{i},i=1,2,...,m\}.
\end{equation}
The number of intersecting points is identified with a topological
number called Linking number $L_{k}$ which can be calculated by
Gauss link integral of lattice gauge potential. The three
dimensional Jordan-Wigner transformation mapping spin to fermions
defined a gauge transformation,
\begin{eqnarray}
&&\sigma^{+}_{\textbf{i}}=Uc^{\dag}_{\textbf{i}},\;\;\sigma^{-}_{\textbf{i}}=c_{\textbf{i}}U^{\dag},\;\;
\sigma^{z}_{\textbf{i}}=c^{\dag}_{\textbf{i}}c_{\textbf{i}}-\frac{1}{2},\nonumber\\
&&U=e^{i\sum_{\textbf{j}\neq\textbf{i}}{j^{\alpha}}\omega(\textbf{j}-\textbf{i})}.
\label{29}
\end{eqnarray}
The gauge potential is introduced as $A_{\mu}=\nabla_{\mu}\omega.$
The corresponding magnetic field\cite{bock} is
\begin{eqnarray}
&&B(i,j,k)=\nabla\times{A(i,j,k)}
\end{eqnarray}
with discrete derivative
\begin{eqnarray}
\partial_{i}A(i,j,k)=A_{i+1,j,k}-A_{i,j,k}.
\end{eqnarray}
The integral of the gauge field along $C_{a\alpha}$ is Wilson loop
$W_{C_{\alpha}}$.

The physicists' approach to the linking number of two loops
$C_{\alpha}$ and $C_{\beta}$ is using analogy with
Amp$\acute{e}$re's law. Gauss link integral on the two entangled
loops does not depend on the physical meaning of vector field
tangent to them, they are valid for any physical vector field. So
we think of loop $C_{\alpha}$ as being a wire carrying an electric
current \emph{I}. The line integral of the generated magnetic
field $B$ around the closed curve is $4\pi{L_{k}}I$, where $L_{k}$
is the linking number which count how many times the current winds
around another closed loop. The linking number is just number of
nontrivial intersecting points between $C_{\alpha}$ and the
minimal surface of $C_{\beta}$. One may also deduce this directly
from the Biot-Savart Law. Therefore we define the topological
correlation as
\begin{equation}\label{St}
\hat{S}^{T}_{\alpha\beta}(t)\;\;\;{\propto}\;\;\;m\;g_{\alpha\beta}\;\;\;{\propto}\;\;\;L_{k}\;g_{\alpha\beta}
\end{equation}
with $L_{k}$ as a topological index. A spin flip at the crossing
point $P_{i}$ excites the two loops at same time, thus they have
overlap in the gauge sectors of Hilbert space. Then we derived a
nontrivial contribution to the multispin correlation function of
two loops. This topological correlation survives local fluctuation
and are valid for thermal states. It should be pointed out here
that some intersecting points would disappear during topological
transformation. Those points do not contribute to the topological
correlation. In the following when we mention intersecting points,
it always means the nontrivial intersecting points which can not
be eliminated by topological transformation.

Knotted Wilson loops in the three dimensional spin models above
also commute with Hamiltonian, they are good quantum numbers. The
topological correlation between linked knots can be described by
topological quantum field theory. Witten found the quantum field
theory to describe a wide class of topological invariant from the
non-abelian Chern-Simons theory\cite{witten}. For a link
containing a collection of knotted, linked loops in three
dimensional manifold $L=\{C_{i},i=1,2,3...\}$, we can compute the
multiloop correlation function between its component $C_{i}$,
\begin{equation}
Z(L)=\langle{C_{i}{\cdots}C_{i}}\rangle=Z{(M)}^{-1}\int[{\mathcal{D}}A]{e^{iS}}\prod_{i=1}^{k}W_{l}(C_{i}),
\end{equation}
where $W_{C}=Tr(Pe^{i\oint{A^{\mu}d\gamma^{\mu}}})$ is the Wilson
loop. $S$ is the non-abelian Chern-Simons action
$S=\frac{k}{4\pi}\int_{M}Tr(A\wedge{dA}+\frac{2}{3}A\wedge{A}\wedge{A})$,
the first term in the non-abelian action represents the local
magnetic Helicity
\begin{equation}
H_{elicity}=\epsilon_{ijk}B_{i}\partial_{j}A_{k}.
\end{equation}
$Z(L)$ is a gauge invariant which provides us a topological
quantity to classify the geometric configuration of linked loops
at different eigenstates.

We compact the three dimensional lattice manifold into $S^3$ in
the continuum limit. A sophisticate link embedded in this $S^{3}$
may be decomposed into two parts by cutting the $S^3$ using an
$S^2$ as section manifold. We choose a proper cutting point so
that the left piece $M_L$ contains most complicated stuff of the
link, and keep the right piece $M_R$ as the simplest link
configuration. The Feynmann path integral on $M_{L}$ determines a
vector $|\chi\rangle$ in the physical Hilbert space, the Feynmann
path integral on $M_{R}$ determines a dual vector $|\psi\rangle$.
The partition function is the natural paring of the two vectors
\begin{equation}
Z(L_{+})=\langle\chi|\psi\rangle,
\end{equation}
$L_{+}$ denote the over crossing of the two loops on the $M_{R}$.
The separated two parts of link puncture four points on the cross
section manifold. The four punctured points represent two spin
fields $\sigma$ and their corresponding antiparticles. Braiding
these quasiparticles on $S^{2}$ leads two different configuration
manifolds of the link, which we denote as $M_{1}$ and $M_{2}$. The
corresponding path integral on $M_{1}$ and $M_{2}$ contributes
another two different vectors $|\psi_{1}\rangle$ and
$|\psi_{2}\rangle$. The two vectors can be viewed as being
generated by applying braid group operations on the original dual
vector $|\psi\rangle$ of $M_{R}$. In topological quantum field
theory, the Skein relation is encoded in
\begin{equation}
{\alpha}\langle\chi|\psi\rangle+\beta\langle\chi|\psi_{1}\rangle+\gamma\langle\chi|\psi_{2}\rangle=0.
\end{equation}
Namely,
\begin{equation}\label{d}
{\alpha}Z(L_{+})+{\beta}Z(L_{0})+{\gamma}Z(L_{-})=0.
\end{equation}
$L_{0}$ denotes the zero crossing in $M_{1}$ and $L_{-}$ denotes
the under crossing in $M_{2}$. Here the three parameters
$\{\alpha,\beta,\gamma\}$ depends on the special models. This
relation is the fundamental constrain on the multi-loop
correlation of sophisticate links in our exact three dimensional
spin models.

The topological correlation is invariant no matter how large one
loop expands continuously. Any local change that break the
topology of a link during evolution would sent out a signal by
correlation function. The linking number $L_{k}$ would jump from
$m$ to $m\pm{i},(i\neq0)$, so does the topological correlation
function $S^{T}\propto{L_{k}}$. If the loop breaks apart at the
crossing lattice site, the topology of link changed, this change
would be detected by the topological correlation function.

As one loop expands without cutting the other loop, the
eigenstates correspondingly evolute from a state of one loop to
another in Hilbert space, but the topological correlation remains
invariant. Therefore local perturbation does not destroy the
topological signal if the two loops does not intersect. This
topological quantum entanglement does not depend on temperature,
so it is a very good candidate for quantum memory. Three
dimensional materials in natural are very promising for
topological quantum computation.

\begin{figure}
\begin{center}\label{knot}
\includegraphics[width=0.3\textwidth]{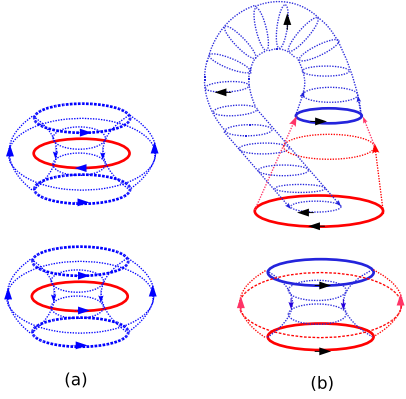}
\caption{(a) one loop wraps another loop, the chirality of the two
loops in the upper is opposite and is same in the lower
half-plane. (b)The exchange of two loops. The upper diagram
depicts the world sheet of two loops with one loop flips its
chirality during the exchange. The lower diagram depicts the case
that two loops keep their chirality invariant during the
exchange.}
\end{center}
\end{figure}

\section{statistics of strings in three dimensions}

The statistics of point particles is characterized by the phase
change of many body wave function during the exchange of
particles, for bosons,
$|\psi(x_{1},x_{2})\rangle=e^{i2\pi}|\psi(x_{2},x_{1})\rangle$,
and for fermions,
$|\psi(x_{1},x_{2})\rangle=e^{i\pi}|\psi(x_{2},x_{1})\rangle$. We
are led back to the initial state
\begin{equation}
|\psi(x_{1},x_{2})\rangle=e^{i2\gamma\pi}|\psi(x_{2},x_{1})\rangle.
\end{equation}
After performing the exchange twice. The anyon statistics falls in
the range of phase $\gamma$ varying continuously from $1/2$ to
$1$. In three dimensions, any wrapping path of one particle around
another can be topologically contracted to a point, that means two
interchange is equivalent to an identical map, so the statistics
of point particles is either fermionic or bosonic.

The emergent string excitations bear exotic statistics in three
dimensions. For the simplest illustration, if two unentangled and
unknotted vortex rings leap through each other, a nontrivial
statistical phase may emerge as Berry's phase\cite{niemi}. The
leapfrogging vortex rings are a vivid demonstration of loop braid
group\cite{baez} for a collection of unknotted, unlinked loops.

When it comes to discrete loops in three dimensional lattice, many
exotic phenomena pop out. Any closed loop in the three dimensional
exactly solvable models discussed above commute with the
corresponding Hamiltonian. The string operator along an oriented
path $C_{\alpha}$ reads
\begin{equation}
\hat{O}_{\alpha}=\prod_{i\in{C_{\alpha}^{\circlearrowleft}}}^{n_{\alpha}}\sigma^{\gamma_{i}}_{i},
\end{equation}
$\circlearrowleft$ denotes the counterclockwise order of operator
product. Its complex conjugate string operator
$\hat{O}_{\alpha}^{\dag}=\prod_{i\in{C^{\circlearrowright}_{\alpha}}}^{n_{\alpha}}\sigma^{\gamma_{i}}_{i}$
has a clockwise product order. Since $[\hat{O}_{\alpha},H]=0$, the
ground state of loop is an eigenstate of $\hat{O}_{\alpha}$
following the equation
\begin{equation}
\hat{O}_{\alpha}|\psi_{\alpha}\rangle=z|\psi_{\alpha}\rangle,z=\pm1.
\end{equation}

In the degenerate Hilbert space of ground state, an initial state
may end up with different final states after adiabatic evolution
along a closed path. The conventional Berry phase needs to be
generalized to a matrix of inner product between degenerate
states. Notice that a loop sweeps out a two dimensional world
sheet, the Berry phase
$\phi=\langle{\psi(R)}|\nabla_{R}|\psi(R)\rangle$ of point
particle along an one dimensional trajectory must be extended to
cover the extra dimension of loop $C_{\alpha}(s)$, where $s$
measures the distance along the curve. We start from the general
eigenfunction of time-dependent Schr$\ddot{o}$dinger equation, an
adiabatic evolution in the degenerate  subspace follows
\begin{equation}\label{berry-a}
|\psi_{\alpha}(C(s))\rangle=e^{-i\int_{0}^{T}E(t)}e^{i\phi_{\alpha\beta}}|{\psi_{\beta}}(C(s))\rangle,
\end{equation}
where
\begin{equation}\label{berry-b}
\phi_{\alpha\beta}=i\int_{0}^{T}dt\int_{T}{{dV}}\langle{\psi_{\alpha}}|{\partial_{t}\psi_{\beta}}\rangle
\end{equation}
is the extended Berry phase. We first consider an unknotted,
unlinked loop $\emph{C}_{\alpha}$. The minimal surface
$\Sigma_{\alpha}(t)$ bounded by $\emph{C}_{\alpha}$ subtends a
solid angle $\Omega_{\alpha}(R,t)$ at point $R$. A collection of
solid angles of different loops
$\{\Omega_{\alpha}(R,t),\Omega_{\beta}(R,t),\cdots\Omega_{\gamma}(R,t)\}$
constitute one geometric representation of loops. In the continuum
limit, the gradient of the solid angle field $-\nabla\Omega(R,t)$
induces an effective magnetic field which circulates around the
loop. This magnetic field could be computed from the Biot-Savart
law. The extended Berry phase in Eq. (\ref{berry-b}) has explicit
expression in terms of solid angle field,
\begin{equation}\label{berry-solid}
\phi_{\alpha\beta}=i\int_{0}^{T}dt\int_{T}{{dV}}\langle{\psi_{\alpha}}|\psi_{\beta}\rangle\partial_{t}\Omega(R,t).
\end{equation}
If we assume the probability density
$\rho_{\alpha\beta}=\langle{\psi_{\alpha}}|\psi_{\beta}\rangle$ is
homogeneous, the quantum mechanical phase is proportional to the
volume which is encapsulated by the world sheet bounded by the two
loops $\emph{C}_{\alpha}$ and $\emph{C}_{\beta}$.

Exotic statistics phase is strongly dependent on the chirality of
loops. For example, consider a red loop and a blue loop with the
same orientation(Fig. 9), the exchange of the two loops sweeps out
a torus surface, the statistical phase is proportional to the
volume of the torus ${V}_{\alpha\beta}^{torus}$. If the chirality
of one loop flips during the exchange, the world sheet becomes
Klein bottle which is unoriented manifold with only one side.
There is no definite volume integral for the quantum mechanical
phase in this case. Here we take another equivalent but simpler
exchange procedure by wrapping the solid red loop around the
dashed blue loop. The dashed blue loop enclosed in the torus
surface of world sheet behaves like electric current which
produces a magnetic field following the Biot-Savart law
\begin{equation}
dB_{b}=c_{0}r^{-3}d\textbf{s}_{b}\times{{\textbf{r}}}.
\end{equation}
The volume integral enclosed by the torus is computed from
\begin{equation}
V={\int}dV={\int}dB_{b}\times{d\textbf{s}_{r}}\cdot{dR}.
\end{equation}
If the two loops have the same chirality, the statistical phase
factor is proportional to the volume of a torus world sheet, it is
positive. On the other hand, if the loops have opposite chirality,
the volume integral is negative. Generally speaking, two loops
with opposite chirality behave like fermions, while two loops with
same chirality obey bosonic statistics.

Another key element governing the statistics of loops is the total
number of particles confined along the loop. The vortex excitation
in the minimal surface surrounded by the loop behaves like
quasiparticles of magnetic flux. We simplify the statistics of two
loops as that between two quasi-particles. If the sum of spins of
one loop is half integer, the quasiparticle is fermion. If the
total spin is integer, the quasiparticle is boson. Moreover,
quasiparticles must carry the orientation of loops, so they have
internal degree of freedom which contributes nontrivial
statistical phase calculated above.

When the loops are knotted and linked, the exchange of two loops
brings topological transformation of configuration of the links.
We first untie, unentangle the two loops to the simplest circle,
and then exchange them. Following an inverse braiding procedure,
we restore the link configuration at their new position, so that
one string evolves exactly into the same configuration of another.
The Reidmeister moves are the fundamental propagators from one
state to another in the Hilbert space of degenerate ground state.
In the three dimensional lattice, the exchange of two knots
follows recursive Skein relation step by step. The statistical
phase factor is no longer a simple complex number but a complex
matrix inherited from the topological constrain.

To operator on quantum states, we need to do braiding operations
on the crossing point of Wilson loops one by one, and deduce the
statistical matrix by using Skein relation. There is a different
way to understand the mathematical braiding from physicist's point
of view. As shown by Witten, the correlation function of Wilson
loop operators in Chern-Simons theory is a topological invariant
of Jones polynomial. We take two neighboring loops
$L_{0}=(\emph{C}_{\alpha},\emph{C}_{\beta})$ as one link
configuration, and do a surgery by cutting one crossing point.
Then we braid the intersecting point on the cross section and glue
them together again, it finally leads to new links, $L_{1}$ and
$L_{2}$. The partition function of the new links and the original
link follows the recursion relation
$AZ(L_{0})+BZ(L_{1})+CZ(L_{2})=0$. We repeat this braiding
operation on all the crossing points until all of the links are
untied, then we exchange them, and braiding them up following an
exact reverse operation. Finally we derive the algebra relation
between statistical matrices. This is the whole strategy to obtain
the statistical matrix algebra of links.

The operations above sound awfully sophisticate, but it is
applicable for both the continuous loop and discrete loops. As for
the discrete loops, we have another different way to their
algebraic statistical relation. Inspired by Levin and Wen's
algebraic statistics relation for point particles hopping on two
dimensional lattice\cite{levin}, we derived a general algebraic
statistic relation of discrete loops using the projection operator
$\hat{T}_{\beta\alpha}=|\psi_{\beta}\rangle\langle\psi_\alpha|$,
which indicates the hopping from loop $\emph{C}_{\alpha}$ to
$\emph{C}_{\beta}$. If we ignore the orientation of the loop and
take the flux quanta surrounded by the loop as one giant
particles, the algebraic statistic relation reads
 \begin{equation}
\hat{T}_{\alpha\beta}\hat{T}_{\gamma\alpha}\hat{T}_{\alpha\lambda}=e^{i\theta}\hat{T}_{\alpha\lambda}\hat{T}_{\gamma\alpha}\hat{T}_{\alpha\beta}.
\end{equation}

The statistic matrix strongly relies on the orientation of loops
and the total number of particles confined along the loops. The
total number of particles is not conserved when the loops
transformed into another loop configuration. The hopping operator
$\hat{T}_{\alpha\beta}$ must be represented by a matrix encoding
the information of both the number of particles and orientation.
Suppose loop $\emph{C}_{\alpha}$ consists of $n$ particles, and
$\emph{C}_{\beta}$ consists of $m$ particles, the hopping operator
between $\emph{C}_{\alpha}$ and $\emph{C}_{\beta}$ is
\begin{equation}\label{Tmatrix}
\hat{T}_{\alpha\beta}=\left(%
\begin{array}{cccc}
  h_{1i} c_{1}^{\dag}c_i & h_{1j}c_{1}^{\dag}c_j & \cdots & h_{1k}c_{1}^{\dag}c_k \\
  h_{2i}c_{2}^{\dag}c_i & h_{2j}c_{2}^{\dag}c_j & \cdots & h_{2k}c_{2}^{\dag}c_k \\
  \vdots & \vdots & \ddots & \vdots \\
  h_{ni}c_{n}^{\dag}c_i & h_{nj}c_{n}^{\dag}c_j & \cdots & h_{nk}c_{n}^{\dag}c_k \\
\end{array}%
\right),
\end{equation}
where $c^\dag_1, \cdots, c^\dag_n$ are the creation operators of
particles along loop $\emph{C}_{\alpha}$ and $c_j~(j=1,\cdots,m)$
are the annihilation operators of particles along loop
$\emph{C}_{\beta}$. In three dimensional spin models, these
fermions are introduced by three dimensional Jordan-Wigner
transformation Eq. (\ref{29}) which maps spins to fermion
operators. The order of indices $(i,j,...,k)$ encodes the
orientation of $\emph{C}_{\beta}$. For a special case that
$\emph{C}_{\alpha}$ and $\emph{C}_{\beta}$ have the same particle
numbers, the hopping matrix is diagonal for loops with the same
orientation and is skew-diagonal for loops with opposite
orientation. The explicit hopping matrix on a torus of world sheet
is
\begin{equation}\label{Tmatrix-torus}
\hat{T}_{torus}=\left(%
\begin{array}{cccc}
  h_{1i} \hat{t}_{11} & 0 & \cdots & 0 \\
  0 & h_{2j}\hat{t}_{22} & \cdots & 0 \\
  \vdots & \vdots & \ddots & \vdots \\
  0 & 0 & \cdots & h_{nk}\hat{t}_{nn} \\
\end{array}%
\right),
\end{equation}
while the hopping matrix on a Klein bottle take the form of
\begin{equation}\label{Tmatrix-klein}
\hat{T}_{Klein}=\left(%
\begin{array}{cccc}
  0  & \cdots & 0 & h_{1n}\hat{t}_{1n} \\
   \vdots& \cdots & h_{2n-1}\hat{t}_{2n-1} & 0 \\
  0 & \cdots & \cdots & \vdots \\
  h_{n1}\hat{t}_{n1} & 0 & \cdots & 0  \\
\end{array}%
\right),
\end{equation}
where $\hat{t}_{ij}=c_{1}^{\dag}c_j$ is the local hopping
operator. Since Wilson loops project out many degenerate ground
states, even if the loops are not entangled or knotted, the
hopping operator does not always commute with each other, i.e.,
\begin{equation}
[\hat{T}_{\alpha\beta},\hat{T}_{\gamma\lambda}]\neq0.
\end{equation}
So there is no longer such a simple statistic algebraic relation
as that for point particles in Ref. \cite{levin}. We have to
develop new algebra on quantum braiding of knotted and linked
loops, this exploration would be presented in some works
elsewhere.

\section{conclusion}

In summary, a class of three dimensional spin liquids have been
constructed and chiral spin liquids can be obtained by following
brief doping rules on conventional three dimensional lattices. For
example, we introduced Kitaev-type 's coupling on the
$Cu$-sublattice of crystal green dioptase
${C{u}}_{6}{S{i}}_{6}{O}_{3}\cdot6{H_{2}O}$, and derived an
exactly solvable spin model. Its exact ground is spin liquid. If
we dope spin on its bonds and assume that there exist only short
range interactions between nearest neighboring doping spins, an
exact chiral spin liquid emerges. The time reversal symmetry is
spontaneously broken at ground state. Linked, knotted loop
excitations are interesting elementary excitations obeying
non-abelian statistics. The statistical phase factor strongly
relies on the chirality of loops as well as the total number of
particles confined along the loop. Two loops with the same
chirality behave like bosons. Two loops with opposite chirality
obey fermionic statistics. Nontrivial multi-loop correlation
measures topological quantum entanglement which is valid under
continuous configuration transformation. Any change of topology of
links make the Linking number jump from one integer to another  so
that the topological correlation function correspondingly change.

\section{Acknowledgment}

This work was supported in part by the national natural science
foundation of China, the national program for basic research of
MOST of China and a fund from CAS. The authors benefited from
participating the program 2007 on Quantum Phases of Matter in
Kavli Institute for Theoretical Physics China.

\end{document}